\documentclass[manuscript,prl]{revtex4}
\usepackage{graphicx,float}

\begin{document}

\title{ 
\Large \bf 
Anomalous diffusion on dynamical networks: 
A model for interacting  epithelial cell migration\\ }

\author{\large \flushleft 
Stefan Thurner$^{a,b}$\footnote[15]{
 {\bf Correspondence should be addressed to:}\\
 Prof. Stefan Thurner, PhD, PhD;  HNO, AKH-Wien, University of Vienna\\
 W\"ahringer G\"urtel 18-20; A-1090 Vienna, Austria\\
 Tel.: ++43 1 40400 2099;  Fax: ++43 1 40400 3332\\
 e-mail: thurner@univie.ac.at\\
}, 
    Nikolaus Wick$^{c,d}$,
    Rudolf Hanel$^{e}$,
    Roland Sedivy$^{c}$,  
%    Ilja Vietor$^{4,6}$
    and Lukas Huber$^{d,f}$} 

\affiliation{  
    $^{a}$ Klinik f\"ur HNO, Universit\"at Wien; Austria,  
    $^{b}$ Institut f\"ur Mathematik, NuHAG, Universit\"at Wien; Austria,
    $^{c}$ Klinisches Institut f\"ur Pathologie, Universit\"at Wien; Austria, 
    $^{d}$ Institute for Molecular Pathology, Vienna; Austria,
    $^{e}$ Institut f\"ur Biomedizinische Technik und Physik, Universit\"at Wien; Austria,
    $^{f}$ Institut f\"ur Anatomie  und Histologie, Universit\"at Innsbruck; Austria 
}

\begin{abstract}
\noindent
We propose a model for cell migration where epithelial cells 
are able to detect trajectories of other cells and try to follow them. 
As cells move along in 2D cell culture, they mark their paths 
by loosing tiny parts of cytoplasm. Any cell moving on a 
surface where other cells have moved before faces a network 
of cell trajectories, which it tries to restrict its motion onto.   
With the Tsallis modification of classical thermodynamics 
one can solve the relevant Fokker-Planck like equation and obtain
experimentally testable distribution functions. 
We compare the model to experimental data  of
normal mammary epithelial cells and cells which have been 
genetically manipulated to change their degree of cell-cell interaction. 
 
\vspace{1cm}

\vspace{0.5cm}
\end{abstract}

\maketitle

\section{Introduction}
A generalization of classical Boltzmann thermodynamics 
as recently introduced by Tsallis \cite{tsallis88,tsallis96,tsallis98}
has offered a beautiful way of describing  the statistics of 
a multitude of physical systems, which generally involve 
some sort of long-range interactions, 
ranging from astrophysics \cite{plastino93,kaniadakis96} 
to low dimensional dissipative chaotic 
systems \cite{lyra98}. Systems described by this thermodynamics   
are often referred to as ''non-extensive''. 
This same modification of classical thermodynamics also 
seems to offer a natural understanding of the subject of 
anomalous diffusion. Anomalous diffusion can be thought of the 
propagation of particles which are restricted to move on a 
certain subspace of full two or three space, such as for  
example the propagation of water molecules in a sponge, or the flow of 
crude oil in porous soil. 
In this work we argue that the motion of living and 
interacting epithelial cells can 
be seen as  a variant of anomalous diffusion on a network which 
changes over time. We support the idea by experimental evidence. 
Our experimental findings moreover strongly support very 
recent work on movements of interacting Hydra cells \cite{upadhyaya01}.  

A living epithelial cell e.g. in two-dimensional cell culture 
is not a static object but moves continuously, a process 
which is called cellular migration \cite{lauffenburger96,lee93}. 
Cell migration is of  vital importance in physiological 
processes such as wound healing or understanding cancerogenesis. 
For example, intact mammary gland tissue is stable during phases 
of milk secretion. 
In residual phases of female estrous cycle, however, it is subject to 
structural remodeling where cells become dynamic entities. 
They proliferate and migrate, eventually forming 
new functional secretory units 
\cite{hennighausen98,robinson95,schedin00}. 
Dysregulation of these events is believed to contribute to cancerogenesis 
\cite{bissell99,foster01,park00}. 
In-vitro experiments preserve important aspects of this epithelial biology  
\cite{aggeler91,fialka96}, 
and migration of single cells has been studied in 
cell culture systems on molecular and mechanical levels 
\cite{chicurel02,lauffenburger96,lee93}. 

Single cells are by no means passive movers and governed by
chance, but influence and control their movements themselves. 
For these movements to occur 
morphological changes are necessary which are realized in   
the formation of distinct subcellular domains. In Fig. 1a a 
single cell is shown during movement. 
After the formation of a so-called lamellipodium, (a V-shaped  
protrusion) this part of the cell attaches to the cell culture surface.  
The cell body is then pulled towards 
the front part (''leading edge'')  via the activation of 
mechanical intracellular linkages. Eventually the ''trailing'' edge, 
a thin, elongated structure at the opposite side of the cell with respect 
to the lamellipodium, 
detaches from the surface which allows a net translocation of the 
cell.  In this ''polarized'' fashion the cell apparently crawls 
on the surface of the cell culture substrate, 
continuously readjusting its direction. During movement cells 
loose tiny parts of their cytoplasm, marking at least parts of their 
trajectories. 
It is known that cells change their movement patterns when the activity of 
proteins involved in cell adhesion is regulated up or down ward. 
When these proteins for example are suppressed, cells 
loose their ability to migrate in a polarized fashion 
\cite{palacios01,hintermann01}. 
As several recent works on velocity distributions of cells show, 
it seems very likely that cell movement depends strongly on the surrounding 
of individual cells. Experimental results range from normal 
diffusion \cite{mombach96} 
to exponential velocity distributions \cite{czirok98}. 
It is the aim of this paper to show that an anomalous diffusion 
approach, already discussed in \cite{upadhyaya01}, might be general 
enough to explain the different experimental results on a common 
theoretical ground. 

\section{Theory}
To describe the dynamics of a single cell which has the ability to detect 
the presence of trajectories of other cells, we propose the following model.
Imagine each cells paints its trail onto the surface. The paint 
slowly fades, and can not be seen after a certain time. 
Another cell which detects the line of paint tries to follow it 
as good as possible. From time to time the cell will be distracted from 
strict path following by random (''thermal'') events.  
At the same time this cell also marks 
its trajectory with paint. As many cells move across the same 
surface there will evolve a network of paths. Since paint fades 
and new trajectories are added to the network, the network 
will undergo slow dynamical changes.  
Any cell equipped with a trail following mode 
will try to restrict its movements onto the network of trails of former cells. 
The situation here is much alike diffusion in porous media   
where particle trajectories 
are restricted onto certain pathways, and can not propagate through 
full two- or three-dimensional space. In the case of cell migration 
an individual cell tries to follow  
the dynamical ''web'' of trajectories other cells have taken before. 

Modeling cell migration as a stochastic diffusion process 
of this kind clearly involves non-linearities in the associated 
Fokker-Planck like equations 
\begin{equation}
\frac{\partial}{\partial t} p({x},t) = 
 -\frac{\partial}{\partial x} \left(F\left[p\right](x,t) p(x,t)\right)
+ \frac{1}{2} \frac{\partial^2}{\partial x^2}\left(D\left[p\right](x,t) p(x,t)
\right)
\quad .
\label{fp}
\end{equation}
To incorporate the idea that cells follow the gradient of metabolites 
emitted by other cells along their trajectories,  
a non-linear flow-term of the form
$F\left[p\right] =\frac{\partial}{\partial x} p$ would be reasonable. 
However, we neglect the flow-term in the following, since 
a posteriori we find  almost no experimental evidence for it. 
To incorporate the anomalous diffusion idea, we decide the functional 
dependence of the diffusion parameter to be $D[p]p = p^{\nu}$. 
With this choice equation (\ref{fp}) can be solved by 
following the entropy approach by Tsallis \cite{tsallis88,tsallis98}. 
Here the the classical entropy definition $S=-\int p(u) \ln p(u) du$
is modified to 
\begin{equation}
S_q=\frac{1-\int  p(u)^q du}{q-1} \quad. 
\end{equation}
By maximizing $S_q$ while keeping energy fixed 
it can be shown that the resulting $p(u)$, i.e. 
\begin{equation}
p_q(x,t)=\frac{\left(1-\beta(t)(1-q)(x-\langle x \rangle(t))^2\right)^{\frac{1}{1-q}}}
 {Z_q(t)}
\end{equation}
also solves Eq. (\ref{fp}) without the flow-term \cite{tsallis96,bologna00}. 
Here $Z_q(t)$ is the generalized $q$ partition function.
The relation of the Tsallis entropy factor $q$, and 
$\nu$ in Eq. (\ref{fp}) is given by $q=2-\nu$ \cite{tsallis96}. 
Note that this distribution is a power law, and only 
in the limit $q \longrightarrow 1$ the classical Gaussian 
$p_1(x,t)= \frac{
\exp^{-\beta(t) \left(x-\langle x \rangle(t)\right) ^2}}{Z_1(t)} $
is recovered. 
It is relatively straight forward to derive the corresponding velocity 
distributions of the associated particles
\begin{equation}
p(v)=B_q\left[1-(1-q)\frac{\beta mv^2}{2}\right]^{\frac{1}{1-q}}
\label{sfp}
\end{equation}
which has been derived  for the first time in \cite{silva98}, 
and which reduces to the usual  
Maxwell-Boltzmann result 
$p(v)=\left( \frac{m}{2\pi k T}\right)^{\frac{3}{2} }
      \exp^{-\frac{\beta mv^2}{2}}$ 
in the limit $q \longrightarrow 1$. Here $\frac{\beta m}{2}$ can be seen 
as an inverse  mobility factor, meaning that large values of $\frac{\beta m}{2}$
are associated with small average cell velocities.   
If cell migration is in fact describable as anomalous diffusion 
governed by a Fokker-Planck equation like Eq. (\ref{fp}) with the 
special choice of $D\left[p\right]$, 
the cell velocity distribution function should follow the form of 
Eq. (\ref{sfp}). 

To further characterize the nature of the diffusion process 
we will also look at normalized temporal autocorrelation functions 
$AC(\tau)= \frac{1}{\sigma^2(\xi)}\langle  \xi(t)\xi(t-\tau) \rangle_t$, 
and at two-point correlation functions 
$C(\tau)=\langle [ \xi(t+\tau) - \xi(t)]^2 \rangle_t$,  
of an underlying process $\xi(t)$,  where  
$\langle.\rangle_t$ denotes the average over time  $t$.  
For (fractional) Brownian motion, the correlator scales 
as $C(\tau)\sim \tau^{2H}$, where $H$ is sometimes called 
the Hurst exponent. For $H=1/2$ the classical 
diffusion result is obtained (Brownian motion), $H>1/2$ indicates 
correlated diffusion, which we expect in our case.

\section{Experimental data}
To record the trajectory of an individual mouse mammary gland 
epithelial cell  
over long time periods of up to 72 hours, 
epithelial cells \cite{data} were seeded 
sparsely onto a 2D plastic culture dish.
The dish was then placed into a mini-chamber 
mounted on an inverted phase contrast video-microscope. 
This provided incubator conditions and periodic 
image acquisition in 90 seconds intervals was performed
\cite{video}. About 900 to 1000 (video) frames were taken 
per measurement. 
For our purposes we only take the first 25 hours of the 
recording since after that cell density becomes high, and 
motion of individual cells restricted, see video
(http://is2.kph.tuwien.ac.at/thurner/CELLMOVE). 
As seen in the videos, cells move within limited regions 
for many timesteps, crossing and following   
its own and other cell's trajectories.  

From the video sequences the motion of individual cells was extracted 
semi-automatically by a commercial motion tracking routine \cite{tracker}. 
Cell nuclei were tracked based on a convolution kernel of a 
100 square pixel field (about half the size of the cell nucleus 
at the given resolution).  
These extracted cell trajectories, see lines in Fig. 1b and c, 
were recorded as the $x_t$ and $y_t$ positions in two dimensions
(the subscript indicating discrete time). 

Due to the limited data length it is more 
useful to study velocity distributions rather than 
spatial distributions directly. 
We work in polar coordinates and transform  
the trajectories $(x_t,y_t)$ to a velocity 
(length-increment $v_x(t)=x_{t}-x_{t-1}$) process 
$v_t = \sqrt{ v_x(t)^2+v_y(t)^2 }$ 
and a direction process 
$\alpha _t= \arctan (x_{t}-x_{t-1},y_{t}-y_{t-1} )$ 
$\in \left[ -\pi, \pi \right]$. 

As a control for the plausibility of our model we genetically modified 
cells from the same cell line in two virus experiments 
\cite{michou99,vietor02,virus}. The basic idea here is to alter  
the cell's motion program such that its interaction with other cells 
and hence its ability to 
follow other cell trajectories becomes effectively reduced. 
In the case of non-interacting cells one  
should expect the model loose its validity, i.e. that cell movement 
would approach classical diffusion, i.e. $q \longrightarrow 1$.   
In the  first experiment, we infected cells with a virus carrying 
the coding sequence of enhanced green fluorescent protein (GFP). 
Since viral infection has been reported to significantly influence 
morphology and molecular homeostasis of mammary epithelial cells
\cite{gaynor85,glotzer00,wolff92,wick02}, 
we expected alterations on cellular migration patterns as well. 
This condition will be referred to 
as ''infected'' in the following. 
In a second experiment we inserted  the coding sequence of  
gene TPA-Inducible Sequence 7 (TIS7) instead of GFP into the cells. 
TIS7, a co-repressor of gene transcription and 
functionally associated with loss of epithelial characteristics 
like cell-to-cell adhesion was chosen to additionally test for eventual 
alterations in cell mobility \cite{vietor02}. 
This experimental condition will be called ''TIS7'' in the 
remainder of the paper and contains consequences 
of the viral infection itself and the effects of over-expressed TIS7.

\section{Results}
For a first demonstration of correlated motion to be present,  
we compute $C(\tau)$ for the individual $x_t$ and $y_t$ components of 
cell trajectories. We find a Hurst exponent of about $H\sim0.7\pm 0.02$ 
for all the experimental conditions (normal, infected, and TIS7), 
indicating correlated motion, see Tab. 1. 

We find the direction process $\alpha_t$ practically equally 
distributed over the interval $\left[ -\pi, \pi \right]$. 
By looking at normalized autocorrelations $AC$ of $\alpha_t$, 
significant (99\% significance level) coefficients 
are found which decay as a power law. 
The situation is shown in Fig. \ref{AC}a for all three experimental 
conditions.  Power decay exponents were fit by a least square routine 
in the region indicated by the 
straight lines; the corresponding values are collected in Tab. 1. 
Errors (standard mean errors over 5 independently moving cells) 
in the fit-regions are somewhat larger than symbol-size and 
were omitted for clarity of the plot. 
Normal cells show lower correlations, 
which decay slightly faster than those of the manipulated cells. 

The normalized autocorrelations for the velocity process $v_t$ 
are given in Fig. \ref{AC}b. Correlation coefficients are statistically 
significant and in the same order of magnitude as the direction series'. 
Again, the tails of the correlations follow a 
power law; for the corresponding exponents, see Tab. 1. 
Here normal cells have higher correlations than infected ones.   
Infected and TIS7 cells show a slower decay than normals. 
Errors here are about twice symbol size. 

In Fig. \ref{Vdist} we show the velocity distribution (histogram) of 
the normal cells in a log-log plot. The solid line is a three parameter 
fit to Eq. \ref{sfp}; fit-parameters for $q$ and $\frac{\beta m}{2}$ for 
all conditions are listed in Tab. 1.  
At the low limit of $v$ the distribution approaches  
a constant, while the right tail clearly decays in a power 
law fashion $ \sim v^{-\gamma}$. 
We find infected and TIS7 cells to move faster on average, as seen in the 
mobility factor or directly in $\langle v_t \rangle$, see Tab. 1. 
 
\section{Discussion}
We provide clear evidence that 
cell migration of interacting cells is a correlated diffusion process. 
This finding is supported by several facts found in our experimental 
data. 
Correlated motion is shown by the 
presence of a Hurst exponent significantly 
larger than 0.5 and by non-vanishing autocorrelation 
functions in the velocity and direction processes. 
This finding is in nice agreement with a recent and similar experiment 
\cite{upadhyaya01}, where an $H$ of $0.62\pm 0.05$ was found. 
The direction autocorrelation functions are found to 
decay in a power like manner,  with exponents consistently smaller 
than 1, opening the possibility for super-diffusion. This obviously 
excludes naive autoregressive or moving average models (ARMA) 
to be realistic candidates for interacting cell migration 
since these models have exponential decays of  
autocorrelation functions. 
This finding means that the direction of cells is 
a ''long memory'' process. 
For velocity autocorrelation functions the power decay 
is smaller but still present for the normal cells. 
The value of the correlation coefficient at $\tau=1$ 
is about the same size as for the $\alpha_t$ process.  

The virus effect is seen in the following way. 
The most obvious difference of normal and manipulated cells 
is seen in their  mobility, which is drastically enhanced 
for manipulated cells. This is expected because we believe that there 
is less effective adhesion proteins present in the infected and 
especially in the TIS7 cells. 
Very similar results have been reported in 
\cite{upadhyaya01}, however, with a different experimental 
strategy: There cells propagated in different aggregates of cell types 
with different adhesion properties to Hydra cells. 

Infected and TIS7 cells show larger correlations in the
direction process, meaning that directions are 
changed less frequently than normal cells do. This 
is expected since these cells detect trails less 
well and lose their ability to follow other cells. 
Once moving they keep   their direction more strictly than 
normal cells. 
For the velocity process we observe the opposite effect: 
normal cells have larger autocorrelations, i.e. normal 
cells change their  velocities less frequently than infected and TIS7 ones. 
We attribute this effect to a more controlled movement of 
normal cells when following a path, or equally, a stronger 
influence of random effects on infected or TIS7 cells with 
damaged movement control. 

We now turn to the main result of this paper. The 
velocity distribution follows the general form of the 
anomalous diffusion model. Maybe most importantly this model  
allows to naturally understand the presence of the power tail 
to the right of the distribution. 
The power can be related (via $q$) to the $\nu$ exponent in 
Eq. (\ref{fp}). For normal cells 
we find $q\sim1.35$ which goes down to $q\sim1.20$ for the infected,  
and to  $q\sim1.11$ for the TIS7 
cells. In the Tsallis thermodynamics framework this means, 
that infected cells approach the classical thermodynamic regime,
while normal cells with their complex motion control system 
are governed by non-extensive statistics. In this framework it is 
clear that for non-interacting cells
ordinary diffusion and Gaussian or exponential 
velocity distributions were observed in \cite{mombach96,czirok98}, 
being extreme cases of the present model. 
If we accept that in our experiments infection 
diminishes the degree of interaction between cells and their 
trajectories, we see that the $q$ value should serve as a sensible 
quantifier of the degree of interaction. 
Again, in \cite{upadhyaya01} very similar behavior 
of 2D moving patterns of Hydra cells were found.  
They reported  $q\sim 1.5$, or a power 
decay in the velocity distributions of about $\gamma\sim 4$ and  
suggest a different way of explaining the movement patterns  based on 
granular flow models. 
Furthermore, these and our findings clearly exclude a Levy flight 
interpretation of cell movements, and led by the observation of 
trail-following tendencies of cells
we believe that cell migration is governed by 
a realization of anomalous diffusion. 

In conclusion, we think that the presented method provides a useful tool to 
qualitatively understand different movement patterns of realistic 
cell migration, and to quantify the degree of interaction by the value of 
the Tsallis $q$.  
Since epithelial cell migration plays such an important role in 
situations like wound healing of skin 
epidermis or cancer metastasis we speculate 
that the above way of quantification could even be a 
practical pathological characterization 
of cells, regarding diverse biological dignities 
and pharmacological responses, in addition to molecular 
and mechanical parameters.  
Finally, we note that the basic idea behind the presented model 
could also be applicable to other fields, like electrical 
flow through networks of biological neurons. There the effective 
network topology is also subject to constant changes due to synaptic 
plasticity (Hebb or other type of learning). 
Electrical current will flow on excitable media restricted 
to a given network and by doing so, will re-shape it by changing 
synaptic weights. Here 
we would predict power law distributions of electrical 
current changes, in  analogy to the $v$ distributions presented here. 
  
\vfill
\noindent 
{\small S.T. would like to thank C. Tsallis and M. Gell-Mann for the 
inspiring workshop on non-extensive thermodynamics they organized 
at the Santa Fe Institute in April 2002. 
We acknowledge M. Cotten for the
preparation of CELO vector stocks. 
This work was supported by Boehringer 
Ingelheim and by the Austrian Science
Foundation grant FWF, P13577-GEN.}
\newpage 

%%%%%%%%%%%%%%%%%%%%%%%%%%%%%%%%%%%%%%%%%%%%%%%%%%%%%%%%%%%%%%%%

\newpage 

\begin{table}[H] 
\begin{tabular}{lccc} 
Measure\hspace{0.5cm} & \hspace{0.5cm} normal cells \hspace{0.5cm} & \hspace{0.5cm} infected \hspace{0.5cm} & \hspace{0.5cm} TIS7 \hspace{0.5cm}   \\
\hline 
$q$                  & 1.350 & 1.201& 1.110 \\
$\frac{\beta m}{2}$  & 1.20  & 0.70 & 0.30   \\
$\langle v_t \rangle$& 38.8 $\mu m/h$ & 51.7 $\mu m/h$ & 71.7 $\mu m/h$ \\
$H$                  & 0.686(24)    & 0.700(18) &  0.739(31)  \\
AC decay exponent: $v_t$        & -0.095       & 0.026     &  -0.009    \\
AC decay exponent: $\alpha_t$& -0.713     &-0.701   &  -0.591    \\
\hline 
\end{tabular}
\caption{Values for the Tsallis $q$, mobility factors 
$\frac{\beta m}{2}$, and average cell velocities $\langle v_t \rangle$, 
for the three cell types. Data was obtained by LS fits to 
the aggregated velocity distributions. 
Additionally, we show mean and standard error 
(from averages over individual trajectories) 
for the Hurst exponent 
and fits to power decay exponents of the tails in the autocorrelation 
functions of the cell velocities $v_t$ and the direction  
process $\alpha_t$.  }
\end{table}

\newpage 

\begin{figure}[H]
\caption{Normal mouse mammary gland epithelial cells migrating on a 
two-dimensional culture surface. 
Large magnification of a single cell (a).  
The leading edge can be recognized as a fan-shaped lamellipodium 
(small arrows) at the bottom of the panel. The trailing edge  
is marked by a star. The direction of migration is indicated by the 
fat arrow.  
%(b) Final image of a representative 
%videomicroscopy experiment. 
%The movement of the center of the nucleus over the past 200 
%time-steps is schematically shown by a black line.  
(b) Extracted trajectory of a single normal cell over 1000 timesteps (25 h). 
} 
\end{figure} 

\begin{figure}[H]
\caption{Autocorrelation function of the direction process $\alpha_t$ 
of cell 
migration for normal, infected, and TIS7 cells (a).  
Infected and TIS7 cells have  somewhat higher values in the 
correlations and  show a softer decay. 
(b) Autocorrelations for the velocity process $v_t$. 
For infected and TIS7 cells the coefficients are lower and 
the curves are nearly constant, 
while normal cells again show a power decay.   
 } 
\label{AC}
\end{figure} 

\begin{figure}[H]
\caption{ 
Cell velocity distribution (histogram) for normal cells. 
For small $v$ the distribution approaches a constant 
%at a value of $v\sim40$ the distribution 
%shows a peak (nor explained by the model), 
and decays for larger $v$ in a clear power law. 
The line represents a fit to the Tsallis velocity distribution. 
The virus experiments are not shown for clarity; their fit 
parameters are gathered, however, in Tab. 1. } 
\label{Vdist}
\end{figure}

%%%%%%%%%%%%%%%%%%%%%%%%%%%%%%%%%%%%%%%%%%%%%%%%%%%%%%%%%%%%%%%%%%%%%
\pagestyle{empty}
\newpage 

\begin{figure}[H]
\begin{center}
\vspace{-1.0cm}
\begin{tabular}{r} 
\includegraphics[width=10.0cm]{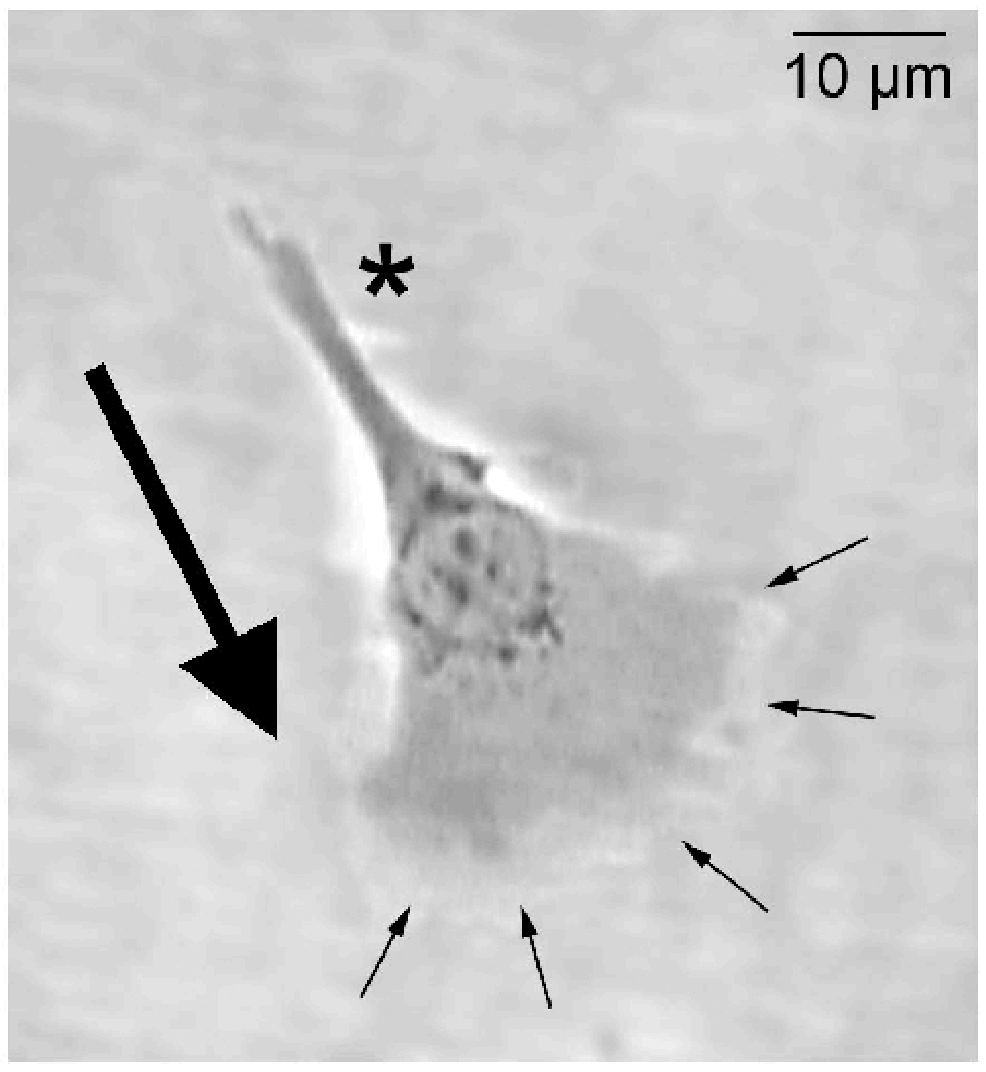} \\
\includegraphics[width=11.2cm]{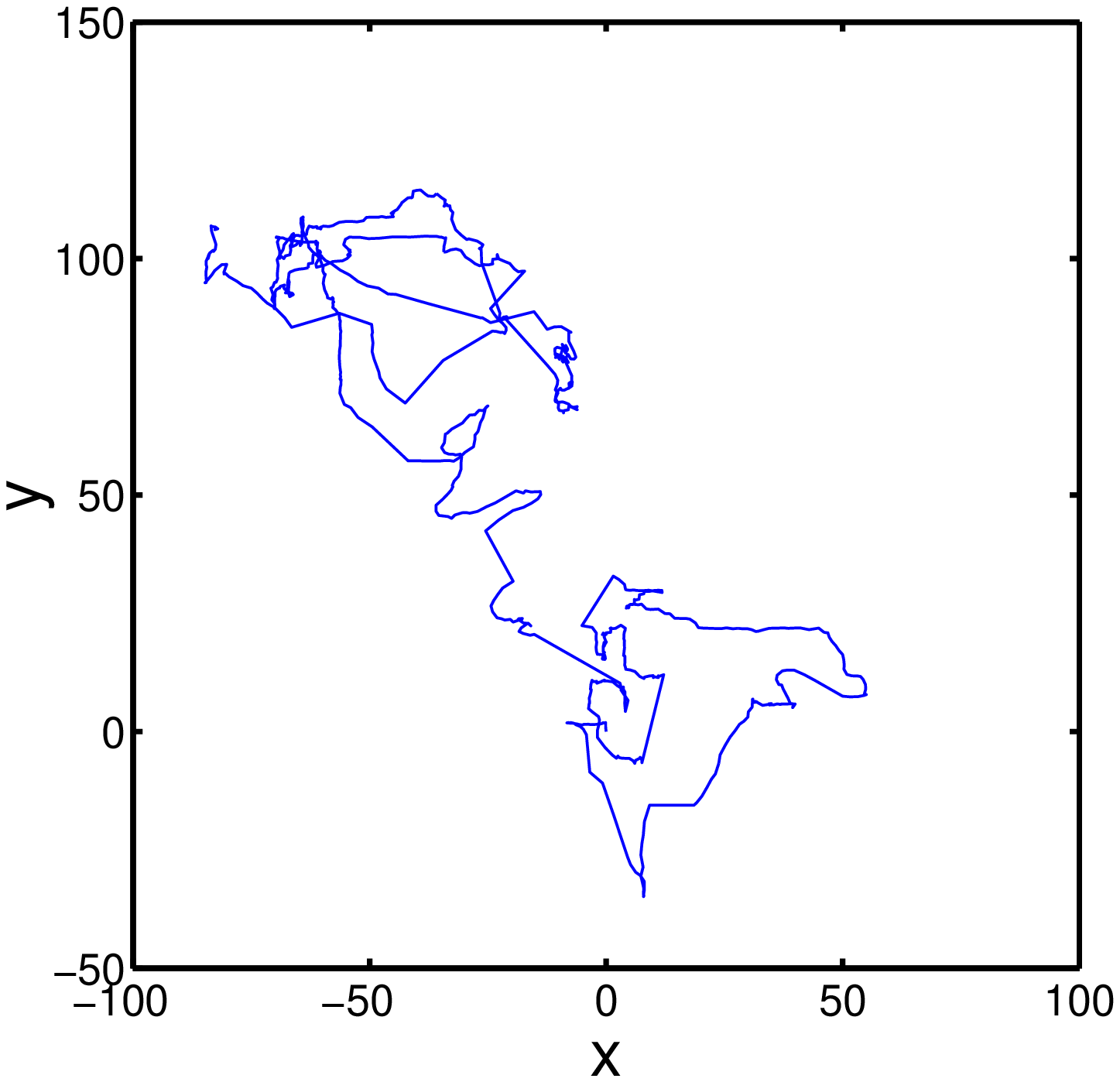} \\
\end{tabular}
\end{center}
\vspace{-14.0cm} {\hspace{11.9cm} \huge (a)}  \\

\vspace{9.1cm}  {\hspace{11.7cm} \huge (b)}  \\
\end{figure} 

\vfill 
\vspace{.5cm} 
\begin{center} 
{\Large FIG. 1 } 
\end{center} 

%\newpage 
%\begin{figure}[H]
%\begin{center}
%\begin{tabular}{r} 
%\includegraphics[width=11.0cm]{fig1c} \\
%\end{tabular}
%\end{center}
%\vspace{-3.5cm} {\hspace{11.5cm} \huge (c)}  \\
%\end{figure} 
%
%\vfill 
%
%\begin{center} 
%{\Large FIG. 1c } 
%\end{center} 
% 

\newpage 

\begin{figure}[H]
\begin{center}
\begin{tabular}{c} 
\includegraphics[width=13.5cm]{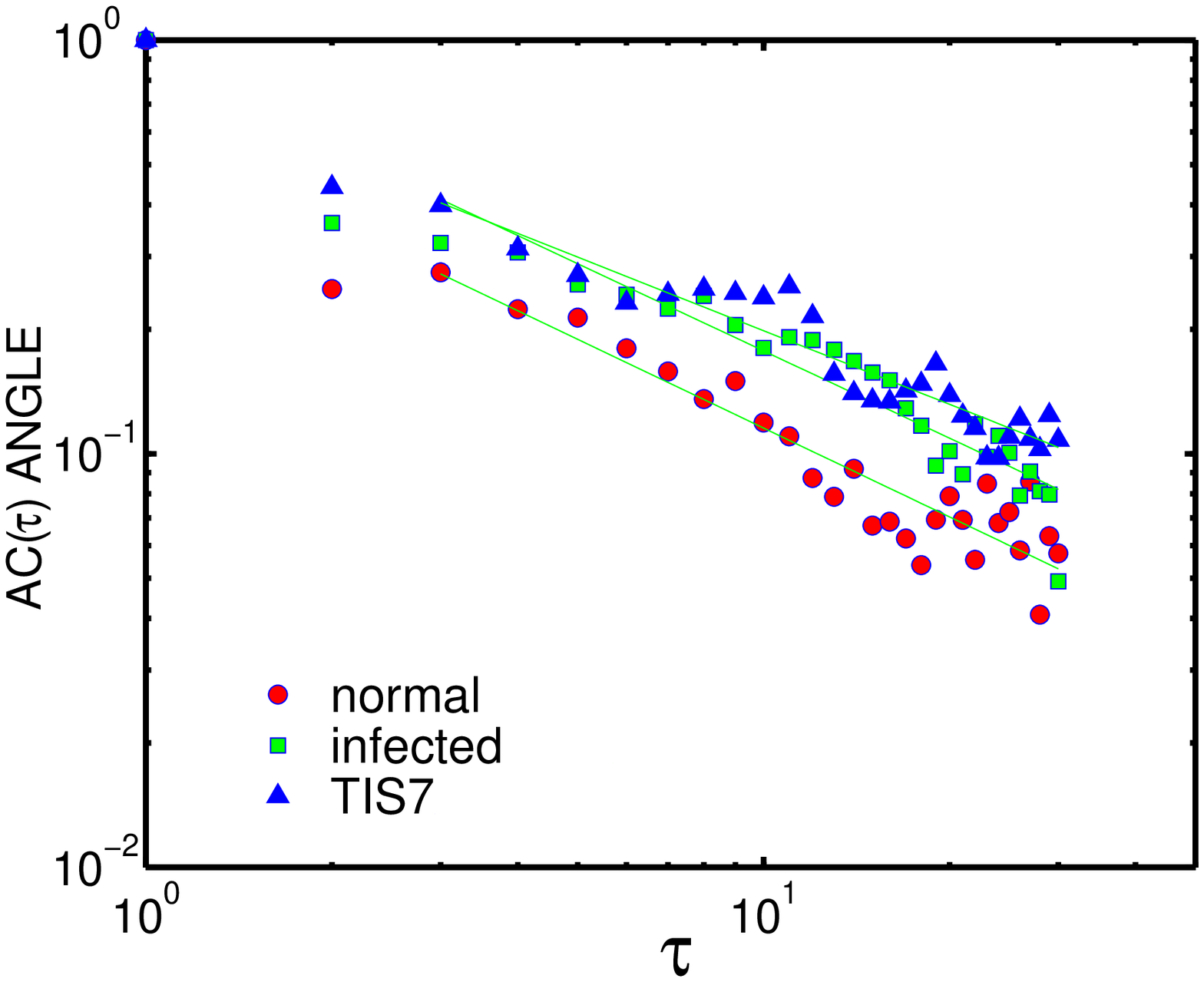} \\
\includegraphics[width=13.5cm]{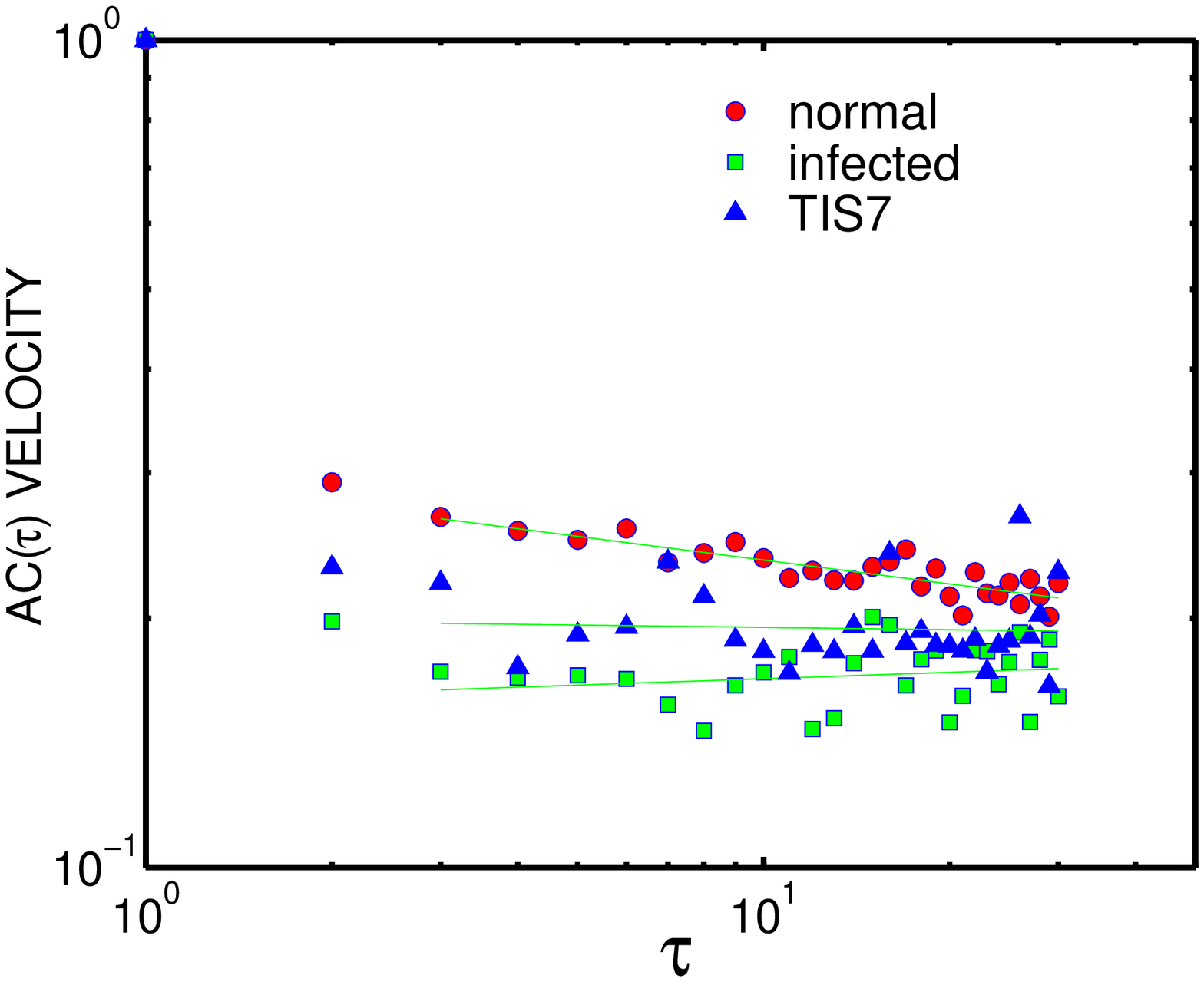} \\
\end{tabular}
\end{center}
\vspace{-14.8cm} {\hspace{13.0cm} \huge (a)}  \\

\vspace{10.0cm}  {\hspace{13.0cm} \huge (b)}  \\
\end{figure} 

\vfill 

\begin{center} 
{\Large FIG. 2} 
\end{center} 

\newpage

\begin{figure}[H]
\begin{center}
\begin{tabular}{c} 
\includegraphics[width=17.5cm]{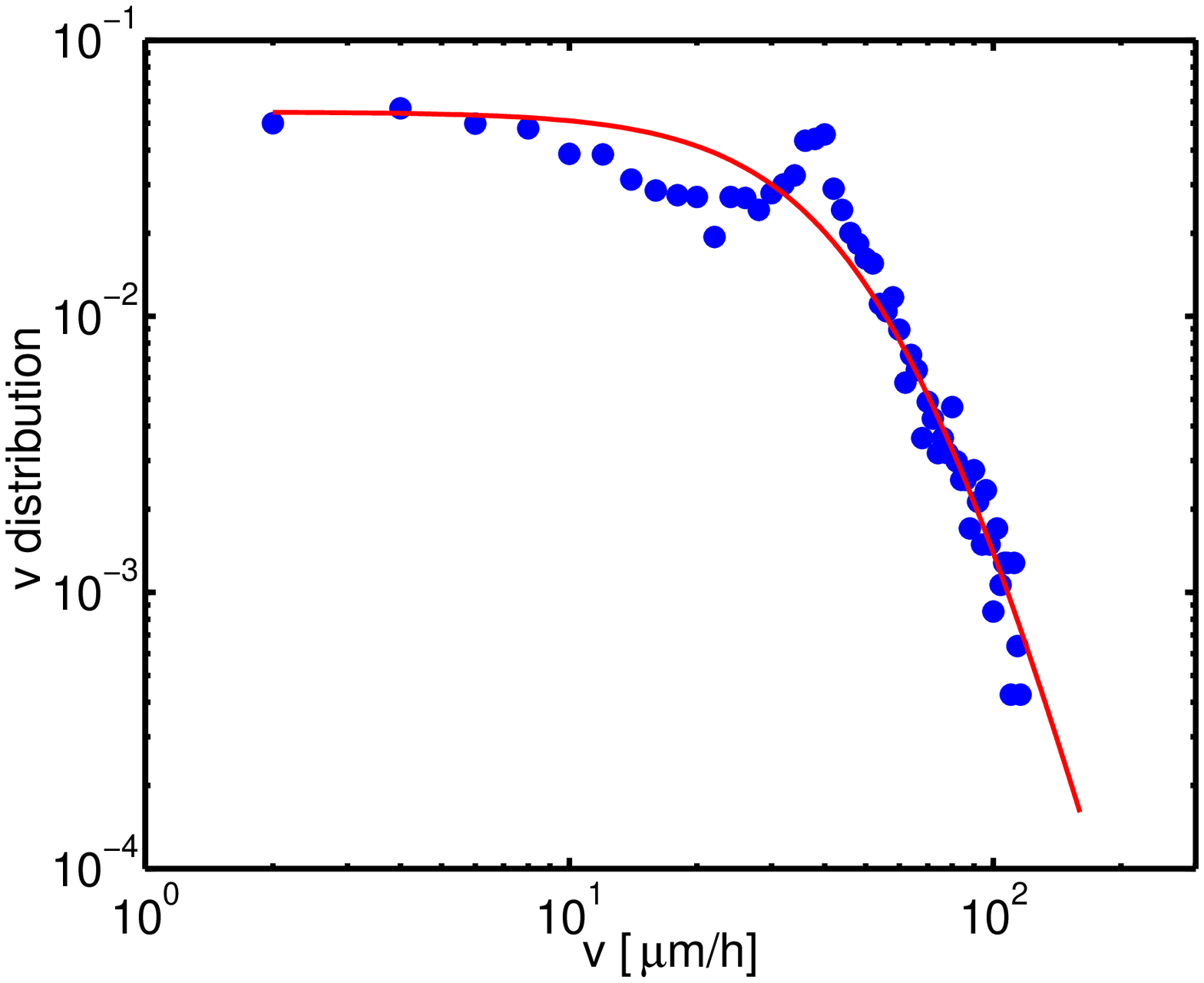} \\
\end{tabular}

\vspace{-3.5cm}
%\hspace{15.3cm} {\huge (c)}  
\end{center}
\end{figure} 

\vfill 

\begin{center} 
{\Large FIG. 3} 
\end{center}


\begin{thebibliography}{99}
\bibitem{virus}
The viral vector derived from the chicken 
embryo lethal orphan and fowl adenovirus type1 and 
has been shown to efficiently serve in overexpression of 
ectopic proteins in epithelial cells \cite{vietor02}.
\end{thebibliography}
\end{document}